\documentclass{osa-article}

\journal{oe}

\newcommand{\abs}[1]{\left\vert #1\right\vert}

\newcommand{\ket}[1]{\left\vert{#1}\right\rangle}

\newcommand{\be}{\begin{equation}}
\newcommand{\ee}{\end{equation}}
\newcommand{\ba}{\begin{array}}
\newcommand{\ea}{\end{array}}
\newcommand{\bqa}{\begin{eqnarray}}
\newcommand{\eqa}{\end{eqnarray}}

\begin{document}

\title{Quantum plasmonic sensing using single photons}

\author{Joong-Sung Lee,\authormark{1} Seung-Jin Yoon,\authormark{1} Hyungju Rah,\authormark{1} \\Mark Tame,\authormark{2,3} Carsten Rockstuhl,\authormark{4,5} Seok Ho Song,\authormark{1} \\Changhyoup Lee,\authormark{4,6}and Kwang-Geol Lee\authormark{1,7}}

\address{
\authormark{1}Department of Physics, Hanyang University, Seoul, 04763, Korea\\
\authormark{2}School of Chemistry and Physics, University of KwaZulu-Natal, Durban 4001, South Africa\\
\authormark{3}National Institute for Theoretical Physics, University of KwaZulu-Natal, Durban 4001, South Africa\\
\authormark{4}Institute of Theoretical Solid State Physics, Karlsruhe Institute of Technology, 76131 Karlsruhe, Germany\\
\authormark{5}Institute of Nanotechnology, Karlsruhe Institute of Technology, 76021 Karlsruhe, Germany\\
\authormark{6}changhyoup.lee@gmail.com\\
\authormark{7}kglee@hanyang.ac.kr
}




\begin{abstract}
Reducing the noise below the shot-noise limit in sensing devices is one of the key promises of quantum technologies. Here, we study quantum plasmonic sensing based on an attenuated total reflection configuration with single photons as input. Our sensor is the Kretschmann configuration with a gold film, and a blood protein in an aqueous solution with different concentrations serves as an analyte. The estimation of the refractive index is performed using heralded single photons. We also determine the estimation error from a statistical analysis over a number of repetitions of identical and independent experiments. We show that the errors of our plasmonic sensor with single photons are below the shot-noise limit even in the presence of various experimental imperfections. Our results demonstrate a practical application of quantum plasmonic sensing is possible given certain improvements are made to the setup investigated, and pave the way for a future generation of quantum plasmonic applications based on similar techniques.
\end{abstract}

\section{Introduction}
Plasmonic effects are successfully exploited in practical photonic sensors, providing much higher sensitivities than conventional photonic sensing platforms~\cite{Homola99a, Lal07, Anker08}. The huge improvement in sensitivity results from the increased optical density of states, given by the strong electromagnetic field enhancement near a metallic surface~\cite{Raether88}. This is linked to the excitation of propagating surface plasmon polaritons (SPPs) at spatially extended interfaces -- hybrid states whose excitation is shared between the electromagnetic field and the charge density oscillation in the metal. The details of the surface plasmon resonances (SPRs) used in photonic sensors and their sensitivity depend on the geometrical and material configuration, forming a large variety of different sensing platforms~\cite{Rothenhausler88, Jorgenson93, Homola99b, Dostalek05, Sepulveda06, Leung07, Svedendahl09, Mayer11}. The most widely used plasmonic sensor is the attenuated total reflection (ATR) setup using the Kretschmann configuration. The simplicity of this configuration has led to its great success in the commercialization of classical biosensing~\cite{Bahadir15}.

The Kretschmann configuration consists of a high index glass material on which a thin metal film is coated. The analyte to be detected is deposited on the other side of the metallic film. The interface is illuminated from the glass side with an incident plane wave in TM polarization (p-polarized) that has a wave vector component parallel to the interface that is larger than the wavenumber in the medium adjacent to the metallic film on the opposite side. The incident field therefore experiences total internal reflection. However, the reflection is attenuated when a propagating SPP is excited at the interface between the metal and the analyte. The excitation conditions depend sensitively on the optical properties of the analyte. Precise sensing in the ATR setup is performed by measuring the variation in either the intensity or the phase of the reflected light at different angles (or different wavelengths) as the refractive index $n_{\text{analyte}}$ of the analyte changes. The measured reflectance curve yields the so-called SPR dip at resonance, across which the phase abruptly changes. These measurements offer a good estimation of the refractive index of the analyte with high sensitivity once the set-up is calibrated. However, the statistical error $\Delta n_{\text{analyte}}$ of the estimation must also be taken into account in evaluating the sensing performance. Most importantly, this error quantifies how precise or reliable the estimated value is. It is known that when the experiment is performed with a classical laser source, the ultimate estimation error, when all technical noises are removed, is inversely proportional to the intensity of the laser light~\cite{Ran06, Piliarik09, Wang11}, i.e., $\Delta n_{\text{analyte}}\propto N^{-1/2}$, where $N$ is the average photon number. This limit is often called the shot-noise limit (SNL) or standard quantum limit. The error, of course, can be reduced by simply increasing the power, but this is not always an acceptable strategy since optical damage might occur when the specimens under investigation are vulnerable~\cite{Neuman99, Peterman03, Taylor15, Taylor16}. Therefore, for sensing in case when photodamage may occur in a low power regime or $N$ is upper-limited by a small number of photons, other strategies have to be put in place in order to go beyond the SNL.

Over the last two decades, the advantages of exploiting quantum resources have been extensively and intensively studied in the field of plasmonics~\cite{Tame13}. Such studies not only provide a better understanding of fundamental quantum plasmonic features, but they also unlock potential applications. One promising application is quantum plasmonic sensing~\cite{Kalashnikov14, Fan15, Pooser15, Lee16, Lee17, Dowran18, Chen18}. In recent years, researchers have introduced quantum techniques using particular quantum states of light for plasmonic sensing in order to beat the SNL in the context of quantum metrology~\cite{Giovannetti04, Boto00, Giovannetti06, Giovannetti11}. Kalashnikov {\it et al.}~experimentally demonstrated the use of frequency-entangled photons in transmission spectroscopy for refractive index sensing in an array of gold nanoparticles with a noise level 70 times lower than the signal~\cite{Kalashnikov14}. Pooser {\it et al.}~also experimentally measured a sensitivity that is $5$~dB better than its classical counterpart by using two-mode intensity squeezed states in the Kretschmann configuration~\cite{Fan15, Pooser15}. Lee {\it et al.}~studied more fundamentally the role of quantum resources combined with plasmonic features in quantum plasmonic sensing and their potential use~\cite{Lee16,Lee17}. Very recently, Dowran {\it et al.}~used bright entangled twin beams to experimentally demonstrate a $56\%$ quantum enhancement in sensitivity, compared to state-of-the-art classical plasmonic sensors~\cite{Dowran18}. Also, Chen {\it et al.}~evaluated the usefulness of their taper-fiber-nanowire coupled system with two-photon plasmonic N00N states for quantum sensing~\cite{Chen18}.

Most of the aforementioned quantum plasmonic sensing schemes rely on transmission or absorption spectroscopy. The change of intensity of the transmitted (or reflected) light after propagation through the sensing platform is analyzed with a variation in sensing samples. It is known that the photon number state $\ket{N}$ is the optimal state for single-mode transmission spectroscopy, leading to a maximal enhancement in precision compared to the classical benchmark~\cite{Monras07, Adesso09, Alipour14, Meda17}. More interestingly, when the state $\ket{N}$ is used, the quantum-to-classical noise ratio for the same average photon number used -- quantifying the amount of quantum enhancement -- does not depend on the photon number $N$, but only on the total transmittance $T_{\text{total}}$. For an absolute comparison with state-of-the-art classical plasmonic sensing, a much higher $N$ photon state is desired to match the high mean photon number of the coherent states used. However, the use of single photons is sufficient to demonstrate the same relative enhancement as obtainable by higher photon number states $\ket{N}$ in transmission spectroscopy.

In this work, we use single photons as inputs in a plasmonic ATR sensor with the Kretschmann configuration. For our sensor to be understood in the context of transmission spectroscopy we treat the actual reflection of a single photon from the ATR setup as a transmission through the ATR setup, as in Ref.~\cite{Lee17}. 
The generation of the single-photon state (the signal) is heralded by a detection of its twin photon (the idler), due to the quantum correlation of photon pairs initially produced via spontaneous parametric down conversion (SPDC). As a sample to analyze, a blood protein in an aqueous solution with different concentrations is chosen. Out of $\nu$ single-photons sent to the ATR setup, we measure the number $N_{\text{t}}$ of transmitted single-photons. We repeat the independent and identical sampling $\mu$ times, assumed to be large enough, to calculate the standard deviation $\langle \Delta N_{\text{t}}\rangle$ where $\langle ..\rangle$ denotes the average over $\mu$ repetitions. These statistical quantities are exploited to quantify the error of estimation in our transmission spectroscopy. We show that the measured estimation errors beat the SNL that would be obtainable by a coherent state of light with the same average photon number as the single photon. Here, the comparison to the SNL is made for the same input power of $N$ and the sampling size of $\nu$, allowing us to focus more on fundamental aspects of using single photons. The quantum enhancement in the error is achieved even in the presence of significant losses, including all experimental imperfections. All of these imperfections diminish the total transmittance $T_{\text{total}}$, subsequently reducing the enhancement. We discuss how the enhancement could be further improved in our setup in a systematic way according to our theoretical analysis, that also explains the experimental results well.

\section{Experimental scheme}

\begin{figure}[h!]
\centering
\includegraphics[width=11cm]{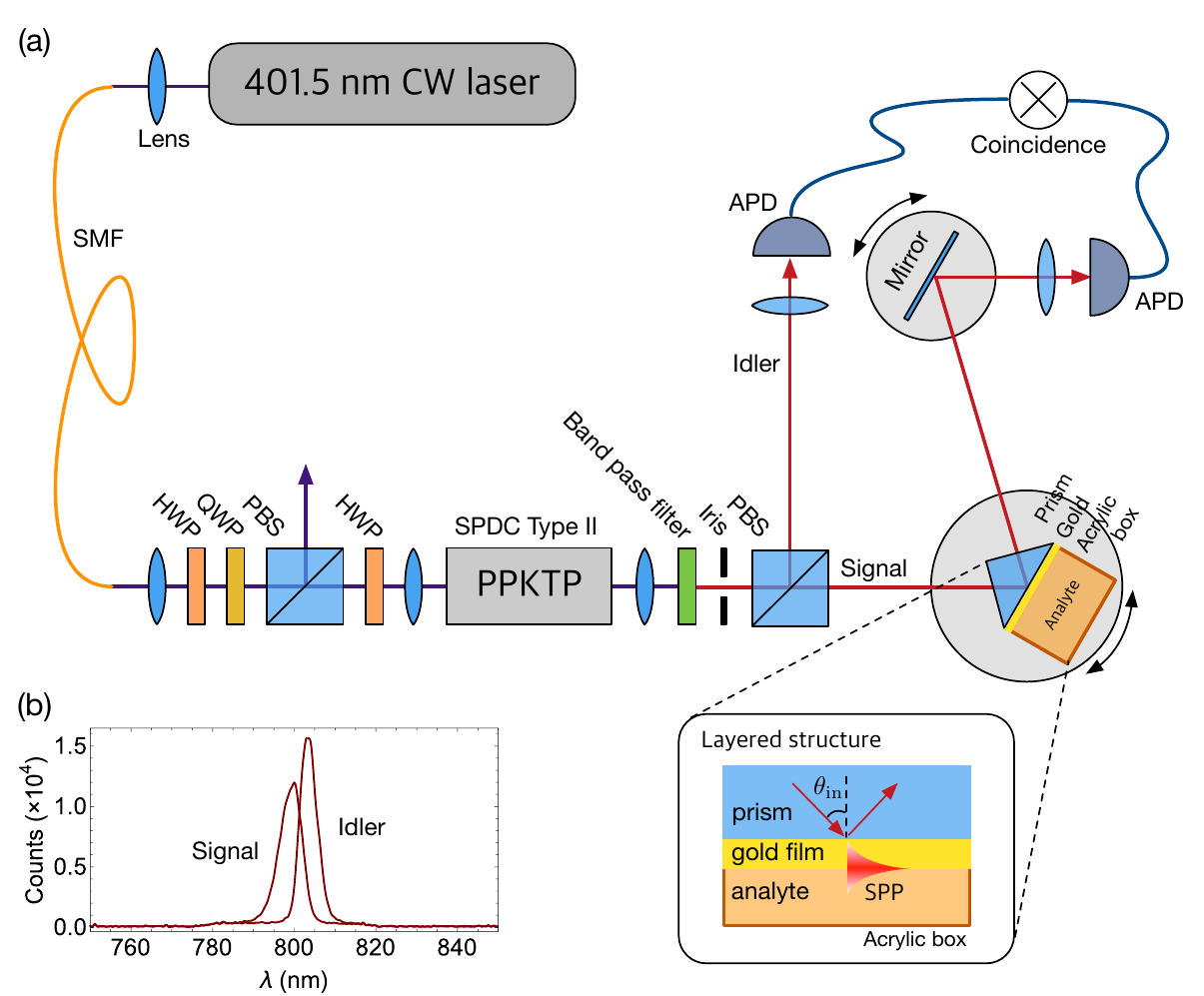}
\caption{
(a) Experimental setup. A continuous wave pump beam at $401.5$~nm is filtered to be a single mode with a particular polarization that maximizes the rate of the photon pair generation through the nonlinear crystal. Such initial filtering is carried out before being injected into the nonlinear crystal (PPKTP). The output beams from the crystal are also filtered via a band pass filter (Thorlabs FBH 800-40) centered at $800$~nm with a width of $40$~nm and then collimated by an iris. Orthogonally polarized pair of photons are split into separate arms through the polarization beam splitter. The photon in the idler mode is directly sent to an APD with a temporal resolution of about $1$~ns, where the detection of a photon heralds the presence of a single photon in the signal mode, which is used as a signal for sensing. This heralded signal photon is sent to the ATR setup, which consists of a prism, a gold layer of about $57$~nm, and an acrylic box that contains the fluidic analyte (see the inset for the layered structure). We then count the number of single photons in the signal mode over the sampling with a size of $\nu=10^4$, conditioned on the cases when a detection event is triggered in the idler mode within the time window of $25$~ns (The time window of the coincidence detection is determined by a FPGA used. The count rate of the idler photon is about $2\times10^{5}$~cps, so that the probability for the twin photons to be detected in the different time windows is nearly zero). We repeat the sampling $\mu=10^3$ times to extract the statistical features of the estimation. 
(b) The spectrum of the output beams are measured by a spectrometer. The central wavelengths of the photon pairs are located at $799.16$~nm and $803.47$~nm with the FWHM of $6.67$~nm and $5.01$~nm, respectively. The wavelengths can be tuned via controlling the temperature of the oven.
}
\label{setup} 
\end{figure}

The schematic of our experiment is shown in Fig.~\ref{setup}(a). A continuous wave diode laser (MDL-III-400, CNI) at $401.5$~nm pumps a nonlinear crystal (periodically poled potassium titanyl phosphate, PPKTP) in a temperature-tunable oven. Its temperature is set to $20^{\circ}$C. It produces pairs of orthogonally polarized photons at $799.16$~nm and $803.47$~nm with a FWHM of $6.67$~nm and $5.01$~nm in the same spatial mode via phase-matching for collinear type-II SPDC. The measured spectra of the generated photon pairs are shown in Fig.~\ref{setup}(b). The produced photons pairs can be approximately written as $\ket{\text{SPDC}}\approx \ket{00}+\epsilon\ket{11}$ with $\epsilon\ll1$. The photon pairs are split into two spatial modes via a polarization beam splitter. One of the photons, the idler photon, is directly sent to an avalanche photodiode single-photon detector (APD, SPCM-AQR-15, PerkinElmer), while the other photon, the signal photon, is fed into the ATR sensing setup. When an idler photon is detected by the APD, it heralds the existence of a twin single photon in the signal mode due to the quantum correlation in photon numbers. In the ATR setup, mounted in a rotation stage for angular modulation, the prism is coated with a gold film of about $57$~nm thickness, where we also install a container made of acrylic glass for a fluidic analyte to be put, as depicted in Fig.~\ref{setup}(a). For an evaluation of our quantum plasmonic sensor, we choose bovine serum albumin (BSA) in aqueous solution with different concentrations \cite{Peters75}. The acrylic container is cleaned by deionized (DI) water before and after measurements for each concentration.

Two kinds of experiments are performed in this work. First, we carry out an incident angular modulation from $66.5^{\circ}$ to $69^{\circ}$ using the heralded single-photon source for BSA concentrations of $0\%$ and $2\%$ as analytes. Here, the concentration $C$ of the BSA is calculated as a ratio of the weight (g) of the BSA powder to $100$~ml of DI water-BSA solution, e.g., $1\%=1~{\text{g}}/100~{\text{ml}}$~\cite{Singh05}. The weight is measured by an electronic scale that has a resolution of $0.01$~g. Second, we measure the change of the transmittance at a fixed incident angle for different concentrations of BSA ranging from $0\%$ to $2\%$ in $0.25\%$ steps.

For each kind of experiment, we post-select the cases when a detection is triggered in the idler mode from the time-tagged table of detections given by a coincidence detection scheme. This constitutes a scheme for a heralded single-photon source. Out of post-selected successive $\nu$ detections in the idler mode (or equivalently out of $\nu$ single-photons sent to the signal mode), we count how many of the transmitted photons are found in the signal mode, yielding the sample mean $T_{\text{total}}=N_{\text{t}}/\nu$. We set the sample size as $\nu=10^4$ in our experiment. The measured transmittance $T_{\text{total}}$ would be accurate with $\nu\rightarrow \infty$, but in reality where $\nu$ is finite, it fluctuates over repetitions of an identical measurement. The amount of fluctuation, the standard deviation (SD) $\langle \Delta T_{\text{total}}\rangle$ of the sample mean in our case, determines the estimation error of transmittance for a given sample of size $\nu$. To measure this quantity experimentally, we repeat the identical experiment $\mu=10^3$ times, which we assume to be large enough to extract statistical features of interest. From $\mu$ samplings with a size of $\nu$, we calculate the SD of $T_{\text{total}}$ as
\begin{align}
\langle \Delta T_{\text{total}}^{\text{(meas)}}\rangle=\sqrt{\frac{1}{\mu}\sum_{j=1}^{\mu} \left(T_{\text{total}}(j)-\langle T_{\text{total}}\rangle\right)^{2}},
\label{Ttotal}
\end{align}
where $T_{\text{total}}(j)$ denotes the transmittance measured in the $j$th sample of size $\nu$ and $\langle T_{\text{total}}\rangle=\sum_{j=1}^{\mu}T_{\text{total}}(j)/\mu$ denotes the mean of the sample mean. To estimate the refractive index $n_{\text{BSA}}$ of the BSA in the ATR setup, a further post-data processing step is required in the distribution of $T_{\text{total}}(j)$, which will be explained in the next section.

\section{Results and discussions} 
We aim to estimate the refractive index $n_{\text{BSA}}$ of the BSA for given concentrations in the ATR setup by fitting our measured data to a well-known formula for the reflectance $R_{\text{sp}}$ of the Kretschmann configuration~\cite{Raether88}. The reflectance is written as
\begin{align}
R_{\text{sp}}=\abs{\frac{e^{ i 2 k_{2} d} r_{23}+ r_{12}}{e^{ i 2 k_{2} d} r_{23}r_{12} + 1}}^{2},
\label{rsp}
\end{align}
where $r_{lm}=\left(\frac{k_{l}}{\varepsilon_{l}} - \frac{k_{m}}{\varepsilon_{m}}\right)\Big/\left(\frac{k_{l}}{\varepsilon_{l}}+\frac{k_{m}}{\varepsilon_{m}}\right)$ for $l,m\in \{1,2,3 \}$, $k_{l}$
denotes the normal-to-surface component of the wave vector in the $l$th layer, $\varepsilon_{l}$ is the respective permittivity, and $d$ is the thickness of the second layer. Here, the first layer is the prism, the second layer is the gold film, and the third layer is the analyte [see the inset in Fig.~\ref{setup}(a)]. The associated quantum theory for the ATR setup has been discussed in Refs.~\cite{Tame08, Ballester09}. What we measure in the experiment is the light reflected from the ATR setup, but we shall regard the reflected light as the transmitted light through the transducer that consists of the ATR setup, as mentioned before. The transmittance being measured in our experiment is the total transmittance $T_{\text{total}}$. This, unfortunately, is not equal to the reflectance $R_{\text{sp}}$ since photon losses can occur before and after the ATR setup. Some losses even depend on the incident angle since the optical paths are not universally aligned for arbitrary incident angles. Therefore, we normalize $T_{\text{total}}$ by the transmittance $T_{\text{total, air}}$ measured for the case of air used as an analyte medium, which is nearly off-resonant from the plasmonic excitation across the entire range of incident angles considered. Then, the normalized transmittance, showing only the transmittance through the prism setup, is given as
\begin{align}
T_{\text{prism}}=\frac{T_{\text{total}}}{\langle T_{\text{total,air}}\rangle},
\label{Tprism}
\end{align}
where the averaged value $\langle T_{\text{total,air}}\rangle=\sum_{j=1}^{\mu}T_{\text{total,air}}(j)/\mu$ is taken into account. Such normalization is expected to remove all unwanted contributions of losses, i.e., $\langle T_{\text{prism}}\rangle \approx R_{\text{sp}}$.

\begin{figure}[h!]
\centering
\includegraphics[width=9cm]{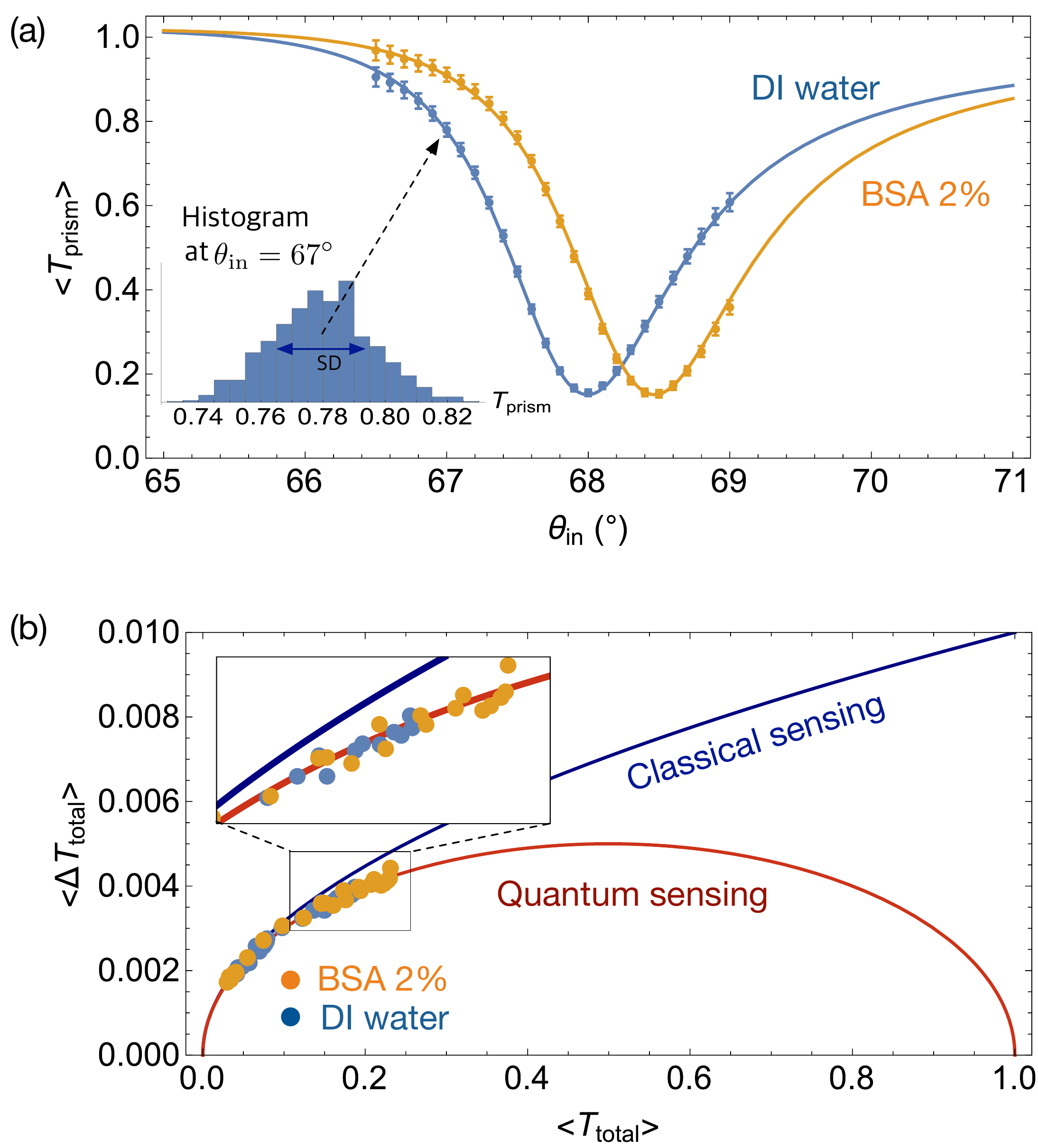}
\caption{
(a) Dots show measured transmittances $\langle T_{\text{prism}}^{\text{(meas)}}\rangle$ of Eq.~(\ref{Tprism}) over the incident angles from $66.5^{\circ}$  to $69^{\circ}$ for BSA concentrations of $0\%$ and $2\%$. Solid lines represent the fitted curves using Eq.~\eqref{rsp}. The error bars are measured as a standard deviation of $T_{\text{prism}}$ in the histogram over $\mu$ repetitions at each incident angle, see the inset for an example. 
(b) The errors $\langle \Delta T_{\text{total}}^{\text{(meas)}}\rangle$, corresponding to the errors $\langle \Delta T_{\text{prism}}^{\text{(meas)}}\rangle$ shown in (a), are represented as a function of $\langle T_{\text{total}}^{\text{(meas)}}\rangle$. The measured errors are compared with theoretically expected errors for classical and quantum sensing when $N=1$, which are given as $\sqrt{T_{\text{total}}^{\text{(true)}}/\nu}$ and $\sqrt{T_{\text{total}}^{\text{(true)}} (1-T_{\text{total}}^{\text{(true)}})/\nu}$, respectively. The comparison  for the same input power of $N$ and the sampling size of $\nu$ clearly demonstrates that the measured errors are below the SNL, defined as the error that would be obtained in classical sensing using a coherent state of light with $N=1$. As the total transmittance of $\langle T_{\text{total}}\rangle$ moves close to zero, the enhancement is not so significant, but the quantum enhancement nevertheless always exists at any value of transmittance.
}
\label{TransmissionNoise} 
\end{figure}

In Fig.~2(a), the measured transmittances $\langle T_{\text{prism}}\rangle$ are shown for the DI water (i.e., $C=0\%$) and the BSA concentration of $2\%$ over the incident angles from $66.5^{\circ}$ to $69.0^{\circ}$. We fit Eq.~\eqref{rsp} to the transmission curves to first obtain the electric permittivity and thickness of the gold film. From a simultaneous fitting to both curves, we obtain $\varepsilon_{\text{gold}}'=-18.2484$ and $\varepsilon_{\text{gold}}''=0.8096$ for the electric permittivity ($\varepsilon_{\text{gold}}=\varepsilon_{\text{gold}}'+i\varepsilon_{\text{gold}}''$) of the gold film at $\lambda=799$~nm, and a thickness of $d=57.41$~nm. Also, the refractive index of the BSA concentration of $2\%$ and the DI water are inferred as $n_{{\text{BSA}},2\%}=1.3325$ and $n_{\text{DI water}}=1.3284$, respectively. The latter is in good agreement with the value ($1.3285$) measured in Ref.~\cite{Daimon07}. 
On the other hand, the error bars are included, obtained as the SD from the histogram of $T_{\text{prism}}$ over $\mu$ repetitions [see the inset in Fig.~\ref{TransmissionNoise}(a), for an example]. It is of great importance to examine if these errors are below the SNL at the same input power considered. To this end, let us consider a coherent state $\ket{\alpha}$ with an average photon number of $N$ and the $N$-photon number state for classical and quantum sensing, respectively. For both cases, we suppose that photon-number-resolving detection is made at the end of the signal channel. When $\mu$ is large enough, it is expected that $\langle\Delta T_{\text{total}}^{\text{(meas)}}\rangle\approx \sqrt{\sigma^{2}/\nu}$, where $\sigma^{2}$ is the variance of the population distribution of the measurement outcomes. Provided that the true value of transmittance is given as $T_{\text{total}}^{\text{(true)}}$, it can be shown that the variances $\sigma^{2}$ are given as $\sigma_{\text{(C)}}^{2} = T_{\text{total}}^{\text{(true)}}N$, and $\sigma_{\text{(Q)}}^{2} = T_{\text{total}}^{\text{(true)}}(1-T_{\text{total}}^{\text{(true)}}) N$, for classical and quantum sensing, respectively~\cite{Loudonbook}. These are given from the fact that the population distributions of the measurement outcomes follow the Poisson and binomial statistics, respectively~\cite{Loudonbook}. In our experiment, $N=1$, for which the APD approximately serves as a photon-number-resolving detector for quantum sensing. The estimator we use is the sample mean, and it is a locally unbiased estimator, so that $\langle T_{\text{total}}^{\text{(meas)}}\rangle = T_{\text{total}}^{\text{(true)}}$. Therefore, the theoretically expected SDs are written as 
\begin{align}
\Delta T_{\text{total}}^{\text{(C)}}&=\sqrt{\frac{T_{\text{total}}^{\text{(true)}}}{\nu}},\label{errorC}\\
\Delta T_{\text{total}}^{\text{(Q)}}&=\sqrt{\frac{T_{\text{total}}^{\text{(true)}}\left(1-T_{\text{total}}^{\text{(true)}}\right)}{\nu}}\label{errorQ},
\end{align}
respectively. The corresponding SDs for the normalized transmittance $T_{\text{prism}}^{\text{(true)}}$ are also given as $\Delta T_{\text{prism}}^{\text{(C)}}=\sqrt{T_{\text{prism}}^{\text{(true)}}/\nu}$ and $\Delta T_{\text{prism}}^{\text{(Q)}}=\sqrt{T_{\text{prism}}^{\text{(true)}}(1-T_{\text{total}}^{\text{(true)}}) /\nu}$, where Eq.~\eqref{Tprism} is taken into account. It is apparent that the noise for classical sensing depends only on the normalized transmittance, whereas the noise for quantum sensing has an additional dependence on the total transmittance. The quantum enhancement can be quantified as a ratio of $\Delta T_{\text{prism}}^{\text{(C)}}$ to $\Delta T_{\text{prism}}^{\text{(Q)}}$, written as
\begin{align}
{\cal R}=\frac{\Delta T_{\text{prism}}^{\text{(C)}}}{\Delta T_{\text{prism}}^{\text{(Q)}}}=\frac{1}{\sqrt{1-T_{\text{total}}^{\text{(true)}}}},
\end{align}
which is always greater than unity. This implies that a quantum enhancement is achieved for any value of $T_{\text{total}}^{\text{(true)}}$. It is interesting that the amount of enhancement is also independent of the average photon number $N$~\cite{Whittaker17}, and the use of a Fock state is always beneficial in reducing the estimation error for any $T_{\text{total}}$ as compared to the classical benchmark. It is evident that the quantum enhancement is truly dependent on the total transmittance $T_{\text{total}}$. Also note that the enhancement is minimal at the resonant point in the SPR curve, where the transmission is attenuated the most, i.e., $T_{\text{prism}}\approx 0$.

\begin{figure}[h!]
\centering
\includegraphics[width=11cm]{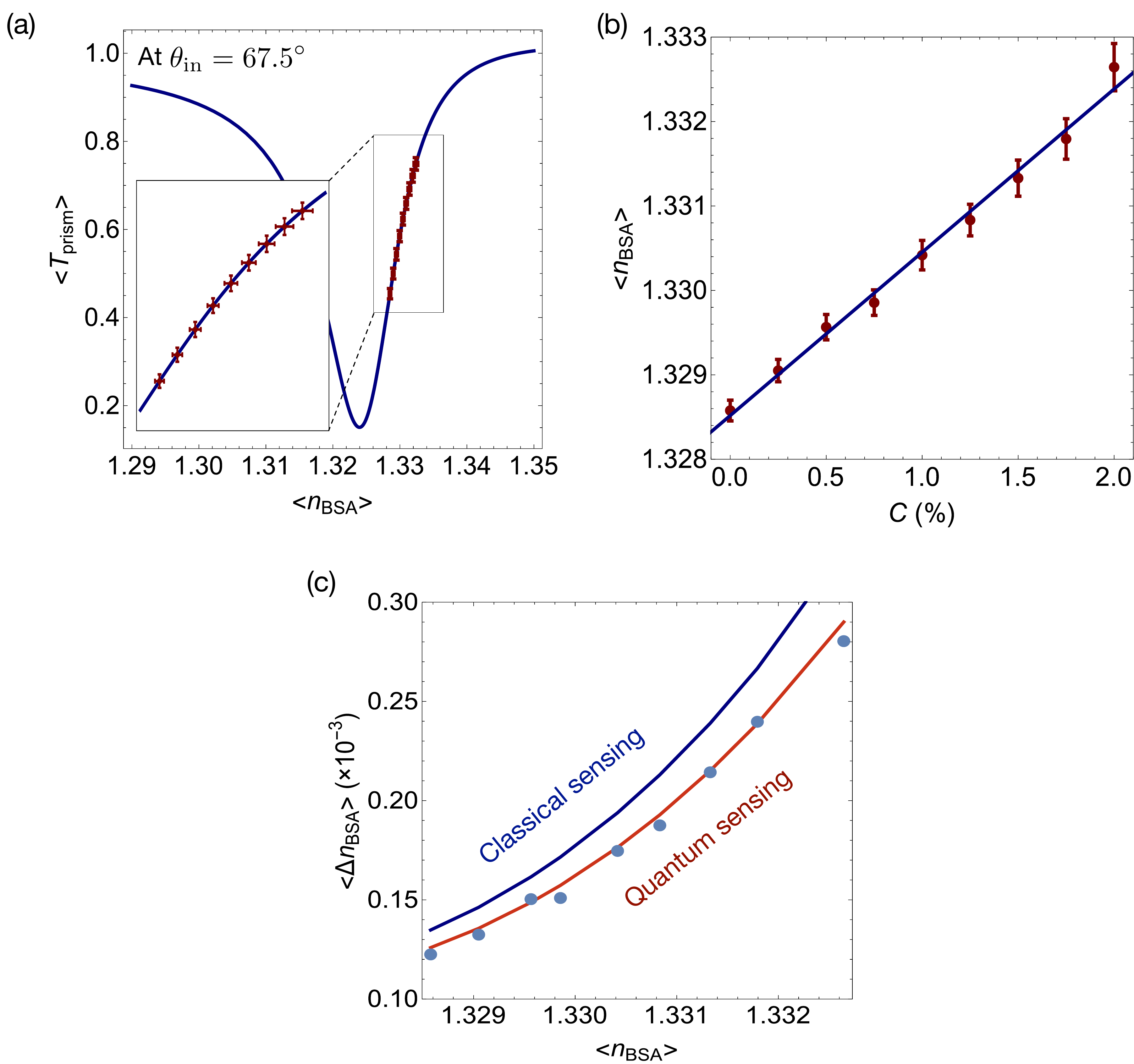}
\caption{
(a) At a fixed incident angle of $\theta_{\text{in}}=67.5^{\circ}$, the transmittance through the prism changes with the refractive index $n_{\text{BSA}}$ of the BSA sample. By measuring the transmittance $\langle T_{\text{prism}}^{\text (meas)}\rangle$, one may infer the refractive index. However, the measured transmittance has a fluctuation represented by $\langle \Delta T_{\text prism}^{\text (meas)}\rangle$, limiting the precision of estimating the refractive index of $n_{\text BSA}$. The dots represent the average of $T_{\text prism}$ and the estimated refractive index $n_{\text BSA}$, whereas the error bars in the horizontal and vertical directions denote the SDs of the histograms for $T_{\text prism}$ and the estimated $n_{\text BSA}$, respectively. Here the BSA concentration varies from $0\%$ to $2\%$ in $0.25\%$ steps. The inset shows a magnified region where the measured data are presented. 
(b) Over $\mu$ repetitions of the experiment, one constructs the histogram of the estimated refractive index $n_{\text BSA}$ for given concentrations. Dots and error bars show the mean and standard deviation of the histogram for the estimated $n_{\text BSA}$ over $\mu$ repetitions, respectively. The solid line represents the averaged dependence of the refractive index with respect to the BSA concentration, yielding the slope of $d\langle n_{\text BSA}\rangle/dC=(1.933\pm0.107)\times10^{-3}$.
(c) The errors taken from (b) are compared with the theoretically expected errors for classical and quantum sensing, each of which is obtained by using the linear error propagation method where $\langle \Delta T_{\text prism}^{\text (meas)}\rangle$ and the derivative of $R_{\text sp}$ with $n_{\text BSA}$ are implemented. The comparison clearly shows that the estimation errors are below the SNL, implying the estimation of the refractive index $n_{\text BSA}$ is more precise when quantum resources are employed .
}
\label{RefractiveIndexNoise} 
\end{figure}

Due to the dependence of the total transmittance on the errors shown in Fig.~\ref{TransmissionNoise}(a), it is more informative to see the experimentally measured total errors as a function of the total transmittance in Fig.~\ref{TransmissionNoise}(b). The errors are compared with the theoretically expected errors of Eqs.~\eqref{errorC} and \eqref{errorQ}. The comparison clearly demonstrates not only that the error bars are in good agreement with quantum theory, but also that they are below the SNL. It is also known that when the population distributions follow the Poisson or binomial distribution, the Fisher information is given as $F=1/\sigma^{2}$. Since the sample mean estimator is locally unbiased, the SD of the histogram is equivalent to the so-called mean-squared-error, which is lower bounded by the Cram\'er-Rao bound~\cite{Cramer46}. The Cram\'er-Rao inequality is written as $\langle \Delta T_{\text total}\rangle\ge (\nu F)^{-1/2}$, where the equality holds only when an optimal estimator is employed. This indicates that the above measured SD can be treated as an ultimate estimate error when photon-number-resolving measurement is considered. 

In the second experiment, we fix the incident angle and vary the BSA concentration from $0\%$ to $2\%$ in $0.25\%$ steps. When the concentration $C$ changes, $T_{\text prism}$ subsequently changes, from which we infer the refractive index of the BSA.
In Fig.~\ref{RefractiveIndexNoise}(a), the relation between the normalized transmittance $T_{\text prism}$ and the refractive index $n_{\text BSA}$ of the sample at an incident angle $\theta_{\text in}=67.5^{\circ}$ is shown (see the solid line) by using Eq.~\eqref{rsp}, with the parameters found from the fitting used in Fig.~\ref{TransmissionNoise}(a). This fitting represents the calibration of the sensor, where the transmittance is linked to a given refractive index. The transmittance $\langle T_{\text prism}\rangle$ for different BSA concentration is measured and the errors are also obtained from the respective histograms [see dots and error bars in Fig.~\ref{RefractiveIndexNoise}(a)]. Due to the fluctuation in the transmittance, one cannot estimate the refractive index with certainty, but rather with a statistical error $\langle \Delta n_{\text BSA}\rangle$, clearly shown in the inset of Fig.~\ref{RefractiveIndexNoise}(a). Including those estimation errors, the measured relation between the refractive index $n_{\text BSA}$ and the BSA concentration $C$ is displayed in Fig.~\ref{RefractiveIndexNoise}(b), where the error bars are obtained from the histogram of the individual estimation of the refractive index over $\mu$ repetitions. The sensitivity of our sensor is calculated as the slope of the linear function that we fit to the experimental data, yielding the slope $d\langle n_{\text BSA}\rangle/dC=(1.933\pm0.107)\times10^{-3}$. Note that the measured sensitivity is in good agreement with the value of $1.82\times 10^{-3}$ previously reported at $\lambda=578$~nm~\cite{Barer54}. We also investigate whether the errors in the estimation of the refractive index are below the SNL for the same input power ($N=1$) considered. We compare the estimation error measured as the SD of the histogram of the estimated refractive indices with the errors calculated using the linear error propagation method \cite{Braunstein94}, written as
\begin{align}
\langle\Delta n_{\text BSA}^{\text (LEPM)}\rangle
=\frac{\langle \Delta T_{\text prism}\rangle}{\abs{\frac{\partial \langle T_{\text prism}\rangle }{\partial n_{\text BSA}}}}.
\label{LEPM}
\end{align}
This method clearly indicates that a high sensitivity provided by plasmonic features is accommodated in the denominator as a derivative of $
\langle T_{\text prism}\rangle$ with $n_{\text BSA}$, whereas the photon number statistics of the input state of light used for sensing is responsible for the numerator $\langle \Delta T_{\text prism}\rangle$. At the incident angle we have chosen, it is clear that the denominator $\abs{\frac{\partial \langle T_{\text prism}\rangle }{\partial n_{\text BSA}}}$ is large when the BSA concentration varies from $0\%$ to $2\%$ [see the slope in Fig.~\ref{RefractiveIndexNoise}(a)]. This part is the same for both classical and quantum sensing, whereas the different photon number statistics leads to a difference in $\langle \Delta T_{\text prism}\rangle$ between classical and quantum sensing. In Fig.~\ref{RefractiveIndexNoise}(c), we compare the experimentally measured $\langle\Delta n_{\text BSA}\rangle$ with the errors $\langle\Delta n_{\text BSA}^{\text (LEPM)}\rangle$, in which the Poisson and binomial statistics are considered for classical and quantum sensing, respectively. It is shown that the estimation error of the refractive index using single photons $\ket{1}$ is lower than that obtainable by a coherent state $\ket{\alpha}$ of light with $\abs{\alpha}^{2}=1$, and in line with that expected from quantum theory.

\section{Discussion}
As before, the quantum-enhancement depends not just on the normalized transmittance $T_{\text prism}$, but rather on the total transmittance $T_{\text total}$. Achieving a larger enhancement requires one to increase the total transmittance as much as possible for a given $\langle T_{\text prism}\rangle$ purely from the sensing prism setup. When the total transmittance is decomposed into successive transmittances as $T_{\text total}=T_{\text before}T_{\text prism}T_{\text after}$, where $T_{\text before}$ and $T_{\text after}$ denote the transmittances before and after the prism setup, we have that the imperfection of the SPDC source reduces $T_{\text before}$, and the finite bandwidth of the source also affects $T_{\text prism}$, while the detection part is responsible for $T_{\text after}$. In our experiment, the APD used has a detection efficiency of $\eta_{\text d}\approx 0.5$ at around $800$~nm, but this could be improved by using a single-photon detector with a higher detection efficiency, e.g., as in Ref.~\cite{Slussarenko17}. In the source part, the broadening of the output spectrum of the generated photon pairs affects $T_{\text prism}$ as it modulates the signal in the SPR curve. Such broadening can be reduced by using a longer nonlinear crystal than the one used in this experiment, which has a length of $10$~mm. Furthermore, the state of photon pairs produced from the SPDC is expected to be $\ket{11}$, upon which the heralding scheme works perfectly, but this is not the case in this experiment since the nonlinear crystal used does not have an anti-reflection coating, and so a reflection of the twin photon can occur even when a photon is found in the idler mode, i.e., the heralded signal state is most likely a mixture of $\ket{1}$ and $\ket{0}$, thus further decreasing $T_{\text before}$. All of these aspects are points of departure for future improvement. Nevertheless, despite all these deficiencies, an improvement in the estimation of the error has been successfully demonstrated by exploiting quantum resources in our plasmonic sensor.

\section{Conclusion}
We have used single photons, known to be optimal states in single-mode transmission spectroscopy, as an input source for a plasmonic sensor using the ATR setup. A quantum enhancement has been observed in a comparison with a classical benchmark obtainable by using a classical state of light with a photon-number-resolving detector. The amount of relative enhancement will be the same even if a higher photon number state $\ket{N}$ is used since it only depends on the total transmittance. We have discussed how our sensing setup could be further improved so as to increase the total transmittance, consequently increasing the quantum enhancement. 

As future work, exploiting two-mode sensing schemes would also help to further increase the quantum enhancement, where a quantum correlation is expected to play a crucial role in enhancing the sensing performance~\cite{Meda17}. One may also consider slightly different sensing platforms that have also been promising for practical purposes, such as using Bloch surface waves in a periodic dielectric stack~\cite{Toma13}, or using a guided mode resonance configuration~\cite{Sahoo17}. Our work indicates that sensing using $N$ photons would be more favorable when the overall transmission is close to unity, i.e., $T_{\text total}\approx 1$, resulting in a much higher enhancement. We believe that our experimental results emphasize the usefulness of single photons or $N$ photons in plasmonic sensing. We hope that this work will help open up future directions in plasmonic sensing, e.g., using sub-Poissonian light sources at a higher optical power regime to directly beat state-of-the-art classical plasmonic sensors.

\section*{Funding}
National Research Foundation of Korea (2016R1A2B4014370, 2014R1A2A1A10050117); Institute for Information \& communications Technology Promotion (IITP-2016-R0992-16-1017).\\

\section*{Acknowledgments}
We thank Jinhyoung Lee for stimulating discussions. This work is supported by the Basic Science Research Program through the National Research Foundation (NRF) of Korea funded by the Ministry of Science, ICT \& Future Planning (MISP), the Information Technology Research Center (ITRC) support program supervised by the Institute for Information \& communications Technology Promotion (IITP), and the South African National Research Foundation and the National Laser Centre.\\


\bibliography{reference}






\end{document}